\documentclass[12pt]{article}

\usepackage{float}
\usepackage{cite}

\usepackage{graphicx}
\usepackage{epsfig}
\usepackage{amssymb}
\usepackage{color}

\def\gtwid{\mathrel{\raise.3ex\hbox{$>$\kern-.75em\lower1ex\hbox{$\sim$}}}}
\def\ltwid{\mathrel{\raise.3ex\hbox{$<$\kern-.75em\lower1ex\hbox{$\sim$}}}}
\def\square{\kern1pt\vbox{\hrule height 1.2pt\hbox{\vrule width 1.2pt\hskip 3pt
   \vbox{\vskip 6pt}\hskip 3pt\vrule width 0.6pt}\hrule height 0.6pt}\kern1pt}
   
\usepackage{amsmath}

\usepackage{cancel}

\allowdisplaybreaks

\begin{document}

\begin{titlepage}

\begin{flushright}
UFIFT-QG-20-05,
CP3-20-38
\end{flushright}

\vspace{0.cm}

\begin{center}
{\bf\large 
One-loop Graviton Corrections to \\ Conformal Scalars on a de Sitter Background
}
\end{center}

\vspace{0.cm}

\begin{center}
D. Glavan$^{1*}$, S. P. Miao$^{2\star}$, T. Prokopec$^{3\dagger}$ and
R. P. Woodard$^{4\ddagger}$
\end{center}
\vspace{0.cm}
\begin{center}
\it{$^{1}$ Centre for Cosmology, Particle Physics and Phenomenology (CP3), \\
Universit\'e catholique de Louvain, Chemin du Cyclotron 2, \\
1348 Louvain-la-Neuve, BELGIUM}
\end{center}
\vspace{-0.3cm}
\begin{center}
\it{$^{2}$ Department of Physics, National Cheng Kung University, \\
No. 1 University Road, Tainan City 70101, TAIWAN}
\end{center}
\vspace{-0.3cm}
\begin{center}
\it{$^{3}$ Institute for Theoretical Physics, Spinoza Institute \& EMME$\Phi$, \\
Utrecht  University, Buys Ballot Building, Princetonplein 5,  \\ 3584 CC  Utrecht, THE NETHERLANDS}
\end{center}
\vspace{-0.3cm}
\begin{center}
\it{$^{4}$ Department of Physics, University of Florida,\\
Gainesville, FL 32611, UNITED STATES}
\end{center}
\vspace{.2cm}
%
We exploit a recent computation of one graviton loop corrections to the
self-mass \cite{Glavan:2020gal} to quantum-correct the field equation 
for a massless, conformally coupled scalar on a de Sitter background.
With the obvious choice for the finite part of the $R^2 \phi^2$ 
counterterm, we find that neither plane wave mode functions nor the 
response to a point source acquires large infrared logarithms. However, 
we do find a decaying logarithmic correction to the mode function and a 
short distance logarithmic running of the potential in addition to the 
power-law effect inherited from flat space.

\begin{flushleft}

\end{flushleft}

\vspace{.0cm}

\begin{flushleft}
$^{*}$ e-mail: drazen.glavan@uclouvain.be \\
$^{\star}$ e-mail: spmiao5@mail.ncku.edu.tw \\
$^{\dagger}$ e-mail: T.Prokopec@uu.nl \\
$^{\ddagger}$ e-mail: woodard@phys.ufl.edu
\end{flushleft}

\end{titlepage}

\section{Introduction}
\label{sec: Introduction}

One of the most challenging problems of inflationary cosmology is 
to reliably quantify the large logarithms that come from graviton
loop corrections. 
This is necessary in order to understand how quantum gravity 
affects matter in inflation.
For example, graviton loop corrections to
the vacuum polarization $i[\mbox{}^{\mu} \Pi^{\nu}](x;x')$ 
change the propagation of dynamical photons, and electromagnetic
forces, through the quantum-corrected Maxwell equation,
\begin{equation}
\partial_{\nu} \Bigl[ \sqrt{-g} \, g^{\nu\rho} g^{\mu\sigma}
F_{\rho \sigma}(x)\Bigr] + \int \!\! d^4x' \, \Bigl[\mbox{}^{\mu}
\Pi^{\nu}\Bigr](x;x') A_{\nu}(x') = J^{\mu}(x) \; , \label{Maxwell}
\end{equation}
where $A_{\mu}(x)$ is the electromagnetic vector potential,
$F_{\rho\sigma} \!\equiv\! \partial_{\rho} A_{\sigma} \!-\! \partial_{\sigma}
A_{\rho}$ is the field strength tensor, $g_{\mu\nu}(x)$ is the 
background metric, and $J^{\mu}(x)$ is the current density. When
equation (\ref{Maxwell}) is solved on a de Sitter background using the 
one graviton loop correction to $i[\mbox{}^{\mu} \Pi^{\nu}](x;x')$ 
in the simplest gauge \cite{Leonard:2012fs}, the electric fields of 
plane wave photons experience a secular enhancement and the Coulomb 
force manifests a logarithmic running \cite{Wang:2014tza,Glavan:2013jca},
\begin{eqnarray}
F_{0i}(t,\vec{x}) 
\!\! & = & \!\!  F^{\rm tree}_{0i}(t,\vec{x}) \Biggl\{ 1 +
\frac{2\hbar G H^2}{\pi} \ln(a) + O(\hbar^2G^2) \Biggr\} \; , \label{photon} \\
\Phi(t,r) \!\! & = & \!\! \frac{Q}{4 \pi a r} 
\Biggl\{ 1 + \frac{\hbar G}{3 \pi a^2 r^2} 
+ \frac{\hbar G H^2}{\pi} \ln(a H r) + O(\hbar ^2G^2) \Biggr\} \; , \label{Coulomb}
\end{eqnarray}
where $G$ is Newton's constant, 
$\hbar$ is the reduced Planck constant,
$H$ is the de Sitter Hubble constant,
and $a(t) \!=\! e^{H t}$ is the de Sitter scale factor. 
The $ \hbar G/(3\pi a^2 r^2)$
correction in (\ref{Coulomb}) is the de Sitter analog of a well-known
flat space result \cite{Radkowski:1970}, but the order $ \hbar G H^2$ logarithms 
in (\ref{photon}) and (\ref{Coulomb}) are new effects due to the 
inflationary expansion of de Sitter. Their physical origin seems to be 
the tendency of redshifting real or virtual photons to acquire momentum 
as they scatter off the continually replenished ensemble of Hubble scale 
gravitons ripped out of the vacuum by inflation. Both effects grow without 
bound in time, and the Coulomb enhancement grows as well at large distances, 
leading to a breakdown of perturbation theory. This raises the fascinating 
possibility of significant loop corrections despite the minuscule quantum 
gravitational loop counting parameter $ \hbar G H^2 \!\sim\! 10^{-11}$. Large 
logarithms have also been found for the field strengths of fermions 
\cite{Miao:2005am,Miao:2006gj,Miao:2012bj} and gravitons \cite{Tsamis:1996qk,
Mora:2013ypa}, and for changes to the background geometry 
\cite{Tsamis:1996qq,Tsamis:1996qm}. It seems inevitable that they occur as 
well in primordial perturbations, which are the principal observables of 
inflation \cite{Weinberg:2005vy,Kahya:2010xh}.

Worries have long been expressed that the large logarithms from loops of
inflationary gravitons might be artifacts of the graviton gauge or poorly chosen 
observables \cite{Unruh:1998ic,Garriga:2007zk,Giddings:2010nc,Urakawa:2010it,
Tanaka:2011aj,Pimentel:2012tw}. There are problems with invoking these 
arguments to deny the possibility of large logarithmic corrections 
\cite{Tsamis:2007is,Miao:2012xc,Basu:2016iua,Basu:2016gyg}, but they do
highlight the importance of correctly computing the numerical coefficients.
This has also been seen directly. Calculations of graviton loop corrections 
on de Sitter background are so difficult that all but one of them have been 
made using the simplest gauge for the graviton propagator \cite{Tsamis:1992xa,
Woodard:2004ut}. However, a heroic computation \cite{Glavan:2015ura} at length 
produced a result for the vacuum polarization in a one-parameter family of 
de Sitter invariant gauges \cite{Miao:2011fc,Kahya:2011sy,Mora:2012zi}. When 
this was used to solve (\ref{Maxwell}) for dynamical photons, a logarithmic 
correction of the same form as (\ref{photon}) was obtained but with a different 
numerical coefficient \cite{Glavan:2016bvp}.

Gauge dependence has long been known to afflict the effective field equations
of flat space \cite{Dolan:1974gu}. Donoghue devised a technique for purging 
it from exchange potentials on a flat space background \cite{Donoghue:1994dn,
BjerrumBohr:2002kt}. One first computes the scattering amplitude for two 
particles that feel the associated force, and then solves the inverse scattering 
problem to reconstruct a gauge-independent potential. Applying this technique 
typically changes numerical coefficients but not the fact of quantum 
gravitational corrections. For example, Bjerrum-Bohr employed Donoghue's 
formalism and found that the simple gauge correction 
of~$\frac{1}{3} \!\times\! \hbar  G/\pi r^2$,
which is evident in expression (\ref{Coulomb}) for $H \!=\! 0$,
becomes $ 6 \! \times\! \hbar G/\pi r^2$ 
in the gauge independent potential~\cite{BjerrumBohr:2002sx}. 

It has recently been understood how to view Donoghue's technique directly as 
a correction to the effective field equations, without going through the 
intermediate step of constructing the $S$-matrix \cite{Miao:2017feh}. This is 
hugely important because it can be applied even to cosmology for which 
the $S$-matrix is not an observable, if it even exists. 
The procedure is to write 
down the position space contributions to the scattering amplitude, and then 
remove the source and observer propagators by applying a series of identities
that Donoghue derived for isolating the leading infrared phenomena 
\cite{Donoghue:1994dn,Donoghue:1996mt}. These identities have the effect of
shrinking higher-point diagrams down to two-point functions which can be viewed
as corrections to the gauge-dependent one-particle-irreducible (1PI) two-point 
functions (such as the vacuum polarization) that appear in the linearized 
effective field equation. However, extending this technique to 
de Sitter will require considerable effort, and it is desirable from both
the conceptual and the practical side to simplify the process as much as possible.

Our program consists of three parts,
\begin{itemize}
\item[(i)]
We first want to identify a simple system that shows large, but possibly 
gauge dependent, logarithms on a de Sitter background. 
\item[(ii)]
Then we will apply a de Sitter space adaptation of the
Donoghue construction in the simple graviton gauge \cite{Tsamis:1992xa,Woodard:2004ut}
to work out reliable coefficients for the large logarithms.
\item[(iii)]
To explicitly 
demonstrate gauge independence, we plan to redo the entire analysis in a 
two-parameter family of generalizations to the simple gauge propagator 
\cite{Glavan:2019msf}. 
\end{itemize} 
One could perform the computation in
a one-parameter family of exact generally covariant gauges~\cite{Mora:2012zi}, 
but that would be needlessly difficult owing to the much more complicated structure of the
propagator. The graviton propagator in a two-parameter family of average generally covariant
gauges has also been worked out~\cite{Frob:2016hkx}, but there seems to be a topological 
obstacle to imposing de Sitter invariant average gauges~\cite{Miao:2009hb}. 
For a discussion on older works on the graviton propagator
see~\cite{Mora:2012zi,Frob:2016hkx} and references therein.

Quantum gravitational corrections to electromagnetism are known to involve 
large logarithms~(\ref{photon}) and~(\ref{Coulomb}) but the intricate analysis we 
intend would be much simpler in a scalar system. The massless, minimally coupled 
scalar suggests itself as a natural choice, and the one graviton loop correction 
to its self-mass has already been derived \cite{Kahya:2007bc}. However, scalar
plane waves are known not to acquire large logarithmic corrections 
\cite{Kahya:2007cm}, and the classical response to a point source is so 
complicated \cite{Glavan:2019yfc,Akhmedov:2010ah} that solving for the one-loop
correction to it might be difficult.

The next most natural candidate is the {\it massless, conformally coupled 
scalar} whose one graviton loop self-mass on a de Sitter background we have recently 
computed \cite{Glavan:2020gal}. 
Note that even though the conformal scalar is insensitive to the cosmological expansion
of the conformally flat de Sitter space, the gravitons running in the loops are not conformally coupled,
and thus mediate the effects of the expansion to the scalar.
Previous works studying graviton loop corrections to conformal 
scalars~\cite{Boran:2014xpa,Boran:2017fsx,Boran:2017cfj} have reported a
correction to the scalar mode function growing faster than the first power of the 
scale factor. This would constitute a huge quantum-gravitational correction, and
investigating its gauge dependence would be of paramount importance.
However, before embarking on the task
of purging gauge dependence,
we set out to check the gauge-fixed computation 
of~\cite{Boran:2014xpa,Boran:2017fsx,Boran:2017cfj} utilizing a simplified
formalism, and here we report no such power-law
enhancement, and no large logarithms, neither for the scalar mode function,
nor for the scalar point source potential, suggesting this system is not
interesting 
for our program.

Some distinction should be drawn between the question of how quantum gravity influences matter in inflation
that concerns us here, and the closely related and equally important question of how quantum matter 
influences gravity in inflation.
In the former case the issue of graviton gauge dependence appears already at leading order as the one-loop
correction to the matter 1PI two-point function is built solely out of graviton propagators. 
On the other hand, in the latter case
this issues never appears at leading order as the one-loop correction to the graviton self-energy is composed solely of
matter fields.\footnote{Strictly speaking this is true for test matter fields, while for matter fields with a classical
condensate the gauge dependence issue at one loop is more complicated.}
Such corrections to gravity from matter loops have been studied for photons~\cite{Wang:2015eaa}, 
as well as for minimally and conformally coupled scalars 
(see~\cite{Park:2011ww,Park:2011kg,Park:2015kua,Frob:2016fcr,Frob:2017smg,Calmet:2015dpa} and references therein).

In this paper we solve the linearized effective 
field equation to check for large logarithms in one-loop corrections to scalar 
plane waves and to the response to a point source. In section~\ref{Effective 
equations of motion} we briefly summarize our result for the self-mass
\cite{Glavan:2020gal}, and use it to quantum-correct the effective field equation.
Sections~\ref{Dynamical scalar} and~\ref{Point source} are devoted to 
perturbatively solving these equations. In section~\ref{Discussion and 
conclusions} we summarize our results and discuss their significance.

\section{Effective equations of motion}
\label{Effective equations of motion}

The tree-level Lagrangian for the system we study in four spacetime dimensions is given by,
\begin{equation}
\mathcal{L} = \frac{R \!-\! 2 \Lambda}{\kappa^2} \sqrt{-g}
	- \frac{1}{2} \partial_\mu \phi \partial_\nu \phi g^{\mu\nu} \sqrt{-g}
	- \frac{1}{12} R \phi^2 \sqrt{-g} \, .
\label{Lagrangian}
\end{equation}
The first of the terms is the Einstein-Hilbert one, where~$\kappa^2\!=\!16\pi G$  
is the gravitational coupling constant, $\Lambda$ is 
the cosmological constant
and~$R$ is the Ricci scalar,
the second is the scalar kinetic term, and the third term
represents the conformal coupling of the scalar to the curvature.
Henceforth, we work in the natural units~$\hbar\!=\!c\!=\!1$,
unless explicitly stated otherwise.
The cubic and quartic interaction vertices between the scalar and the graviton 
are defined by expanding the metric around de Sitter space,
\begin{equation}
g_{\mu\nu} = a^2 
\bigl( \eta_{\mu\nu} + \kappa h_{\mu\nu} \bigr) \, ,
\end{equation}
where~$a(\eta)\!=\!-1/(H\eta)$ is the scale factor given in conformal time coordinate~$\eta$,
the constant Hubble expansion
rate is denoted by~$H$,
and~$h_{\mu\nu}$ is the (conformally rescaled) 
graviton field.
Renormalizing one-loop corrections requires counterterms 
not already contained 
in~(\ref{Lagrangian}). Apart from absorbing divergences originating
from interactions~\cite{Glavan:2020gal},
they also produce a finite local contribution to the one-loop effective action,
\begin{eqnarray}
&&
\Delta\mathcal{L}^{\rm loc.} = 
	\kappa^2 \Biggl\{
	- \frac{\alpha}{2} \biggl[ \square \phi \!-\! \frac{R\phi}{6} \biggr]^2 \sqrt{-g}
	- \frac{\beta}{24} \biggl[ \square \phi \!-\! \frac{R\phi}{6} \biggr] \phi R \sqrt{-g} 
\nonumber
\\
&&
\hspace{3.cm}
	-\, \frac{\gamma}{24} \partial_i \phi \partial_j \phi g^{ij} \phi R \sqrt{-g} 
	- \frac{\delta}{288} \phi^2 R^2 \sqrt{-g}
	\Biggr\} \, . \qquad
\label{L ctm}
\end{eqnarray}

The quantum corrections to the classical behavior of the conformal scalar in de Sitter
are captured by effective field equations, which are most conveniently written for a 
conformally rescaled field,~$\widetilde{\phi} \!=\! a\phi$,
since at tree level~$\widetilde{\phi}$ behaves as a scalar in flat space,
\begin{equation}
\partial^2 \widetilde{\phi}(x)
	- \int \! d^{4}x' \, \widetilde{M}^2_R(x;x') \widetilde{\phi}(x')
	= \widetilde{J}(x) \, .
\end{equation}
Here~$\partial^2 \!=\! -\partial_0^2 \!+\! \nabla^2$ is the 
flat space d'Alembertian 
operator,~$\widetilde{J} \!=\! a^3 J$ is the 
conformally rescaled classical source, and~$\widetilde{M}_R^2$ is the
conformally rescaled renormalized
self-mass-squared, $\widetilde{M}_R^2(x;x') \!=\! (aa')^{-1}\!\times\! M_R^2(x;x')$.
The retarded self-mass corresponds to the sum of the~$(++)$ and~$(+-)$ components
that appear in the Schwinger-Keldysh formalism for nonequilibrium quantum field 
theory~\cite{Schwinger:1960qe,Mahanthappa:1962ex,Bakshi:1962dv,Bakshi:1963bn,Keldysh:1964ud,Chou:1984es,Jordan:1986ug,Calzetta:1986ey}, 
\begin{equation}
\widetilde{M}^2_{\rm R}(x;x')
	= 
	\widetilde{M}^2_{\scriptscriptstyle ++}(x;x')
	+
	\widetilde{M}^2_{\scriptscriptstyle +-}(x;x') \, .
\label{retarded self-mass}
\end{equation}
%
\begin{figure}[H]
\centering
\vskip-0.6cm
\includegraphics[width=13cm]{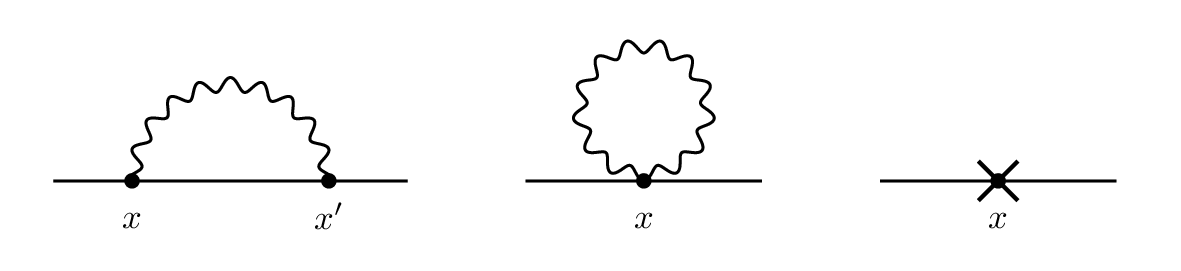}
\vskip-0.7cm
\caption{
One-particle-irreducible diagrams contributing to the scalar self-mass-squared at 
the one-loop order. The solid lines stand for the scalar and wavy ones for 
the graviton. The rightmost diagram  stands for the counterterms.}
\label{diagrams}
\end{figure}
In Ref.~\cite{Glavan:2020gal}
we reported the~$(++)$ component of the renormalized one-loop self-mass,
which receives contributions from diagrams in Fig.~\ref{diagrams}, 
\begin{eqnarray}
&& \hspace{-0.7cm}
- i \widetilde{M}^2_{\scriptscriptstyle ++}(x;x')
	=
	\kappa^2 \partial^2 \partial'^2
	\biggl\{ \biggl[ \frac{\ln(aa')}{96 \pi^2} - \alpha_{\rm} \biggr] \frac{i \delta^4(x\!-\!x')}{aa'} \biggr\}
\nonumber \\
&& 
	+\, \kappa^2 H^2 \partial \!\cdot\! \partial'
	\biggl\{ \biggl[ \frac{19 \ln(aa')}{96 \pi^2} + \beta_{\rm } \biggr] i \delta^4(x\!-\!x') \biggr\}
\nonumber \\
&& 
	-\, \kappa^2 H^2 \vec{\nabla} \!\cdot\! \vec{\nabla}'
	\biggl\{ \biggl[ \frac{5 \ln(aa')}{16 \pi^2} 
	+ \gamma_{\rm } \biggr] i \delta^4(x\!-\!x') \biggr\}
	- \delta_{\rm } \kappa^2 H^4 a^2 \, i \delta^4(x\!-\!x')
\nonumber \\
&& 
	+\, 
	\frac{\kappa^2 \partial^2 \partial'^2}{384 \pi^4} \biggl(
	\frac{1}{aa'} \partial^2 \biggl[ 
		\frac{\ln(\mu^2 \Delta x_{\scriptscriptstyle ++}^2)}{\Delta x_{\scriptscriptstyle ++}^2} \biggr]
	\biggr)
	- \frac{\kappa^2 H^2 (19 \partial^4 \!\!-\! 18 \nabla^2 \partial^2 ) }{ 384\pi^4 }
		\biggl[ \frac{\ln(\mu^2 \Delta x_{\scriptscriptstyle ++}^2)}{\Delta x_{\scriptscriptstyle ++}^2} \biggr]
\nonumber \\
&& 
	+\, \frac{\kappa^2 H^2 \partial^2 \nabla^2}{16 \pi^4}
		\biggl[ 
		\frac{ \frac{1}{2} \ln(\frac{1}{4}H^2 \Delta x_{\scriptscriptstyle ++}^2) + 1}{\Delta x_{\scriptscriptstyle ++}^2} 
			\biggr]
\, ,
\label{M++}
\end{eqnarray}
where the Lorentz-invariant distance squared is
\begin{equation}
\Delta x^2_{\scriptscriptstyle ++}
	= \bigl\| \vec{x} \!-\! \vec{x}^{\,\prime} \bigr\|^2
		- \bigl( | \eta \!-\! \eta' | \!-\! i \varepsilon \bigr)^2 \, ,
\end{equation}
and the physical significance of 
the coupling 
constants~$\alpha, \beta, \gamma, \delta$ can be 
inferred from Eq.~(\ref{L ctm}).
%
%
The ($+-$) component is obtained from the~$(++)$ one by
(i) dropping all the local terms, (ii) substituting all~$\Delta x_{\scriptscriptstyle ++}$'s by
\begin{equation}
\Delta x^2_{\scriptscriptstyle +-}
	= \bigl\| \vec{x} \!-\! \vec{x}^{\,\prime} \bigr\|^2
		- \bigl( \eta \!-\! \eta' \!+\! i \varepsilon \bigr)^2 \, ,
\end{equation}
and (iii) appending an overall minus sign,
\begin{eqnarray}
&& \hspace{-0.7cm}
- i \widetilde{M}^2_{\scriptscriptstyle +-}(x;x')
	=
	-\frac{\kappa^2 \partial^2 \partial'^2}{384 \pi^4} \biggl(
	\frac{1}{aa'} \partial^2 \biggl[ 
		\frac{\ln(\mu^2 \Delta x_{\scriptscriptstyle +-}^2)}{\Delta x_{\scriptscriptstyle +-}^2} \biggr]
	\biggr)
\label{M+-}
\\
&& \hspace{-0.7cm}
	+ \frac{\kappa^2 H^2 (19 \partial^4 \!-\! 18 \nabla^2 \partial^2 ) }{ 384\pi^4 }
		\biggl[ \frac{\ln(\mu^2 \Delta x_{\scriptscriptstyle +-}^2)}{\Delta x_{\scriptscriptstyle +-}^2} \biggr]
	- \frac{\kappa^2 H^2 \partial^2 \nabla^2}{16 \pi^4}
		\biggl[ \frac{ \frac{1}{2} \ln(\frac{1}{4}H^2 \Delta x_{\scriptscriptstyle +-}^2) + 1}
				{\Delta x_{\scriptscriptstyle +-}^2} \biggr] \, .
\nonumber 
\end{eqnarray}
When adding~(\ref{M++}) and~(\ref{M+-}) we make use of the two identities
(that can be found in, {\it e.g.},~\cite{Duffy:2005ue}),
\begin{eqnarray}
&&
\hspace{-0.7cm}
\frac{1}{\Delta x_{\scriptscriptstyle++}^2} 
	- \frac{1}{\Delta x_{\scriptscriptstyle +-}^2}
	=
	\frac{ i \pi }{2} \partial^2 \theta\bigl( \Delta\eta \!-\! \| \Delta\vec{x} \| \bigr) \, ,
\\
&&	
\hspace{-0.7cm}
\frac{\ln \bigl( \mu^2 \Delta x_{\scriptscriptstyle ++}^2 \bigr) }{\Delta x_{\scriptscriptstyle ++}^2}
	- \frac{\ln \bigl( \mu^2 \Delta x_{\scriptscriptstyle +-}^2 \bigr) }{\Delta x_{\scriptscriptstyle +-}^2}
\nonumber \\
&&	\hspace{1.5cm}
	=
	\frac{i \pi}{2} \partial^2 \biggl\{
	\theta\bigl( \Delta\eta \!-\! \| \Delta\vec{x} \| \bigr)
	\biggl( \ln\Bigl[ \mu^2 \bigl( \Delta\eta^2 \!-\! \| \Delta\vec{x} \|^2 \bigr) \Bigr] - 1 \biggr)
	\biggr\}
	\, ,	\qquad
\end{eqnarray}
where~$\Delta\vec{x} \!=\! \vec{x} \!-\! \vec{x}^{\,\prime}$ and~$\Delta\eta\!=\!\eta \!-\! \eta'$,
to form the retarded self-energy appearing in the effective field equations,
\begin{eqnarray}
&& \hspace{-0.7cm}
\widetilde{M}^2_{R}(x;x')
	=
	-\kappa^2 \partial^2 \partial'^2
	\biggl\{ \biggl[ \frac{\ln(aa')}{96 \pi^2} - \alpha_{\rm } \biggr] \frac{\delta^4(x\!-\!x')}{aa'} \biggr\}
\nonumber \\
&& 
	-\, \kappa^2 H^2 \partial \!\cdot\! \partial'
	\biggl\{ \biggl[ \frac{19 \ln(aa')}{96 \pi^2} + \beta_{\rm } \biggr] \delta^4(x\!-\!x') \biggr\}
\nonumber \\
&& 
	+\, \kappa^2 H^2 \vec{\nabla} \!\cdot\! \vec{\nabla}'
	\biggl\{ \biggl[ \frac{5 \ln(aa')}{16 \pi^2} 
	      + \gamma_{\rm } \biggr] \delta^4(x\!-\!x') \biggr\}
	+ \delta_{\rm } \kappa^2 H^4 a^2 \, \delta^4(x\!-\!x')
\nonumber \\
&& 
	-\, 
	\frac{\kappa^2 \partial^2 \partial'^2}{768 \pi^3} \biggl\{
	\frac{1}{aa'} \partial^4 \biggl[
	\theta\bigl( \Delta\eta \!-\! \| \Delta\vec{x} \| \bigr)
	\biggl( \ln\Bigl[ \mu^2 \bigl( \Delta\eta^2 \!-\! \| \Delta\vec{x} \|^2 \bigr) \Bigr] \!-\! 1 \biggr)
	\biggr]
	\biggr\}
\nonumber \\
&&
	+\, \frac{\kappa^2 H^2 (19 \partial^2 \!-\! 18 \nabla^2)\partial^4 }{768\pi^3 }
	\biggl\{
	\theta\bigl( \Delta\eta \!-\! \| \Delta\vec{x} \| \bigr)
	\biggl( \ln\Bigl[ \mu^2 \bigl( \Delta\eta^2 \!-\! \| \Delta\vec{x} \|^2 \bigr) \Bigr] \!-\! 1 \biggr)
	\biggr\}
\nonumber \\
&& 
	-\,\frac{\kappa^2 H^2 \partial^4 \nabla^2}{64 \pi^3}
	\biggl\{
	\theta\bigl( \Delta\eta \!-\! \| \Delta\vec{x} \| \bigr)
	\biggl( \ln\Bigl[ \tfrac{1}{4} H^2 \bigl( \Delta\eta^2 \!-\! \| \Delta\vec{x} \|^2 \bigr) \Bigr] \!+\! 1 \biggr)
	\biggr\}
\, .
\label{retarded self mass}
\end{eqnarray}
The first four terms containing a delta function we refer to as {\it local terms},
while the remaining three terms have support away from coincidence, and we refer to them
as {\it nonlocal terms}.

\medskip

The two physical systems we are interested in are the dynamical scalar 
where~$\widetilde{J}(x)\!=\!0$, and the point source~$\widetilde{J}(x) \!=\! \delta^3(\vec{x})$.
Quantum effects will modify the classical behavior. We have the self-mass-squared 
computed at one loop, so it only makes sense to compute the first correction to the
scalar mode function,
\begin{eqnarray}
\hspace{-0.7cm}
\widetilde{J}(\eta,\vec{x}) = 0
	\ &\Longrightarrow& \
	\widetilde{\phi}(\eta,\vec{x})
	=
	\biggl[ u_0(\eta,k) + \kappa^2 u_1(\eta,k) 
		+ \mathcal{O}(\kappa^4) \biggr] e^{i\vec{k}\cdot\vec{x}} \, ,
\\
\hspace{-0.7cm}
\widetilde{J}(\eta,\vec{x}) = \delta^3(\vec{x})
	\ &\Longrightarrow& \
	\widetilde{\phi}(\eta,\vec{x})
	=
	\frac{- 1}{4 \pi \| \vec{x} \|}
		\biggl[ 1 + \kappa^2 \Phi_1(\eta, \| \vec{x} \|)
			+ \mathcal{O}(\kappa^4) \biggr]  \, ,
\quad
\label{pt source pert expansion}
\end{eqnarray}
where~$u_0(\eta,k)\!=\!e^{-ik\eta}$ is the tree-level mode function
of the monochromatic conformally rescaled field.
Solving for the quantum corrections amounts to solving,
\begin{eqnarray}
- \kappa^2 \bigl( \partial_0^2 \!+\! k^2 \bigr) u_1(\eta,k)
	\!\! &=& \!\! e^{- i \vec{k} \cdot \vec{x} } \int \! d^4x' \, \widetilde{M}_R^2(x;x') \,
		e^{-ik\eta' + i \vec{k}\cdot\vec{x}{\,}' } \, ,
\label{plane wave correction}
\\
\kappa^2 \partial^2 \biggl[ \frac{ \Phi_1(\eta,\| \vec{x} \|) }{ \| \vec{x} \| } \biggr]
	\!\! &=& \!\! \int \! d^4x' \, \widetilde{M}_R^2(x;x') \,
		\frac{1}{ \| \vec{x}^{\,\prime}  \|} \, .
\label{point source correction}
\end{eqnarray}
We solve these two equations in the two following sections, using the one-loop
retarded self-mass from Eq.~(\ref{retarded self mass}).

\section{Dynamical scalar}
\label{Dynamical scalar}

In this section we solve Eq.~(\ref{plane wave correction}) to determine the 
one-loop graviton correction to the conformal scalar mode function at late times
for which~$a\!\to\!\infty$.
It is convenient to split the source on the right-hand side into seven pieces,
\begin{equation}
- \bigl( \partial_0^2 + k^2 \bigr) u_1(\eta,k)
	=
\sum_{n=1}^{7} I_n(\eta,k) \, ,
\end{equation}
where each of them corresponds to one term in the 
retarded one-loop self-mass~(\ref{retarded self mass}),
\begin{eqnarray}
&&
\hspace{-0.7cm}
I_1 =
	- \int\! d^4x' \, \partial^2 \partial'^2 \biggl\{
		\biggl[ \frac{\ln(aa')}{96 \pi^2} \!-\! \alpha_{\rm } \biggr]
			\frac{ \delta^4(x\!-\!x')}{aa'} \biggr\}
     e^{-ik\eta' - i \vec{k} \cdot (\vec{x}-\vec{x}{\,}' )} 
				\, ,
\label{I1def}
\\
&&
\hspace{-0.7cm}
I_2 =
	- \int\! d^4x' \, H^2 \partial \!\cdot\! \partial' \biggl\{
		\biggl[ \frac{19 \ln(aa')}{96 \pi^2} \!+\! \beta_{\rm } \biggr]
			\delta^4(x\!-\!x') \biggr\}
				e^{-ik\eta' - i \vec{k} \cdot (\vec{x}-\vec{x}{\,}' ) } 
				 \, ,
\label{I2def}
\\
&&
\hspace{-0.7cm}
I_3 =
	\int\! d^4x' \, H^2 \vec{\nabla} \!\cdot\! \vec{\nabla}' \biggl\{
	\biggl[ \frac{5 \ln(aa')}{16 \pi^2} \!+\! \gamma_{\rm } \biggr]
			\delta^4(x\!-\!x') \biggr\}
              e^{-ik\eta' - i \vec{k} \cdot (\vec{x}-\vec{x}{\,}' )} 
				\, , \qquad
\label{I3def}
\\
&&
\hspace{-0.7cm}
I_4 =
	\int\! d^4x' \, \delta_{\rm } H^4 a^2
		 \delta^4(x\!-\!x') \, 
       e^{-ik\eta' - i \vec{k} \cdot (\vec{x}-\vec{x}{\,}' )}  
\, ,
\label{I4def}
\\
&&
\hspace{-0.7cm}
I_5 =
	- \frac{1}{768 \pi^3}
	\int\! d^4x' \, \partial^2 \partial'^2 \biggl\{
	\frac{1}{aa'} \partial^4 \biggl[
	\theta\bigl( \Delta\eta \!-\! \| \Delta\vec{x} \| \bigr) 
\nonumber \\
&&	\hspace{2.5cm}
	\times
	\biggl( \ln\Bigl[ \mu^2 \bigl( \Delta\eta^2 \!-\! \| \Delta\vec{x} \|^2 \bigr) \Bigr] \!-\! 1 \biggr)
	\biggr]
	\biggr\}
	e^{-ik\eta' - i \vec{k} \cdot (\vec{x}-\vec{x}{\,}' )} 
\, , \qquad
\label{I5def}
\\
&&
\hspace{-0.7cm}
I_6 =
	\frac{1 }{ 768\pi^3 }
	\int\! d^4 x' \, H^2 \bigl(19 \partial^2 \!-\! 18 \nabla^2
	\bigr) \partial^2 \partial'^2
	\biggl\{
	\theta\bigl( \Delta\eta \!-\! \| \Delta\vec{x} \| \bigr) 
\nonumber \\
&&	\hspace{2.5cm}
	\times
	\biggl( \ln\Bigl[ \mu^2 \bigl( \Delta\eta^2 \!-\! \| \Delta\vec{x} \|^2 \bigr) \Bigr] \!-\! 1 \biggr)
	\biggr\}
	e^{-ik\eta' - i \vec{k} \cdot (\vec{x}-\vec{x}{\,}' )} 
\, ,
\label{I6def}
\\
&&
\hspace{-0.7cm}
I_7 =
	- \frac{ 1 }{64 \pi^3}
	\int\! d^4 x' \, H^2 \nabla^2 \partial^2 \partial'^2
	\biggl\{
	\theta\bigl( \Delta\eta \!-\! \| \Delta\vec{x} \| \bigr) 
\nonumber \\
&&	\hspace{2.5cm}
	\times
	\biggl( \ln\Bigl[ \tfrac{1}{4} H^2 \bigl( \Delta\eta^2 \!-\! \| \Delta\vec{x} \|^2 \bigr) \Bigr] \!+\! 1 \biggr)
	\biggr\}
	e^{-ik\eta' - i \vec{k} \cdot (\vec{x}-\vec{x}{\,}' )} 
\, .
\label{I7def}
\end{eqnarray}
Note that in the last two sources for convenience we have turned 
one~$\partial^2$ into~$\partial'^2$,
as it acts on a function of relative coordinates only.
The first four sources, descending from the local terms in the self-mass, are
straightforward to evaluate,
\begin{eqnarray}
&&
I_1 
= 0 \, ,
\label{I1sol}
\\
&&
I_2 =
	\frac{19}{48\pi^2} (i k H^3 a) \times u_0(\eta,k)
	\, ,
\label{I2sol}
\\
&&
I_3 =
	\biggl[ \frac{5 \ln(a)}{8 \pi^2} \!+\! \gamma_{\rm } \biggr]H^2 k^2\times u_0(\eta,k)
	\, ,
\label{I3sol}
\\
&&
I_4 =
	\delta \, H^4 a^2 
		\times u_0(\eta,k)
		\, .
\label{I4sol}
\end{eqnarray}
The remaining three sources, corresponding to nonlocal terms in the self-mass,
can only produce terms of the form of initial state corrections that decay in time.
This is seen by integrating by parts~$\partial'^2$ onto the classical mode function,
which annihilates it. The only contributions then come from the surface terms
evaluated at the initial time surface, which decay at late times,
\begin{equation}
I_5 = I_6 = I_7 = 0 \, .
\label{zero integrals}
\end{equation}
The contributions from the initial time surface
that we have dropped one should be able to absorb into initial state corrections,
in a manner analogous to what was done 
in Ref.~\cite{Kahya:2009sz},
and are thus not dynamical effects we are interested in.
They can be evaluated as was done
in, {\it e.g.},~\cite{Duffy:2005ue}. 

The three nonvanishing sources~(\ref{I2sol})--(\ref{I4sol}) are all proportional to~$u_0$,
so it makes sense to look for the late time solution for~$u_1$ in the form,
\begin{equation}
u_1(\eta,k) = H^2 f(\eta,k) \times u_0(\eta,k) \,, 
\end{equation}
so that~$f(\eta,k)$ satisfies,
\begin{equation}
\partial_0 \bigl( \partial_0 \!-\! 2ik \bigr) f(\eta,k)
	= - \delta H^2 a^2
		- \frac{19 i k H }{48\pi^2} a
		- \frac{5 k^2 }{8 \pi^2} \ln(a) - \gamma k^2 \, .
\end{equation}
Integrating once produces,
\begin{eqnarray}
&&
\bigl( \partial_0 \!-\! 2ik \bigr) f(\eta,k)
	= - \delta H a
		- \frac{19 i k }{48\pi^2} \ln(a)
		+ \frac{5 k^2 \ln(a)}{8\pi^2 H a}
\qquad
\nonumber \\
&&	\hspace{6cm}
		+ \biggl( \frac{5}{8\pi^2} \!+\! \gamma \biggr)
			\frac{k^2}{H a} 
		+ C(k) ,
\end{eqnarray}
where~$C(k)$ is an integration constant dependent on initial conditions. 
Inverting this first order differential equation
is now straightforward,
\begin{eqnarray}
f(\eta,k) \xrightarrow{a\to\infty}
	- \delta \, \ln(a)
	+ \overline{C}(k)
+ \frac{ik}{H}
	\biggl( \frac{19}{48\pi^2} + 2\delta \biggr) \frac{\ln(a)}{a}
	+ \mathcal{O}(1/a) \, ,
\label{f result}
\end{eqnarray}
where
\begin{equation}
u(\eta,k) = u_0(\eta,k) \times \Bigl[ 1 + (\kappa H)^2 f(\eta,k) \Bigr] \, .
\end{equation}
The first and the third terms in~(\ref{f result}) contain logarithms and
represent unambiguous dynamical effects from graviton loops in de Sitter,
and these are the corrections we are interested in. The second term
in~(\ref{f result}), on the other hand, does not represent a dynamical
correction, but rather can be absorbed into perturbative non-Gaussian
corrections of the initial state, much as in Ref.~\cite{Kahya:2009sz}.

\section{Point source}
\label{Point source}

\begin{figure}[H]
\centering
\includegraphics[width=11cm]{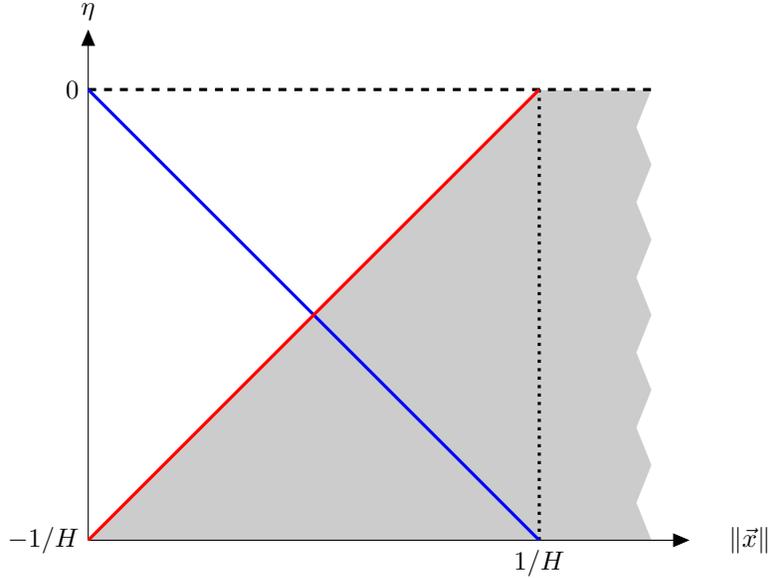}
\vspace{-1cm}
\caption{Conformal diagram of the cosmological patch
of de Sitter space. The system is released at time~$\eta_0\!=\!-1/H$,
with a scalar point source at the origin~$\vec{x}\!=\!0$. The asymptotic future 
corresponds to the~$\eta\!=\!0$ slice. 
The red line denotes the light cone 
of the point source given by~$(\eta\!-\!\eta_0)\!-\! \| \vec{x} \|\!=\!0$, 
while the blue line denotes the  
Hubble distance from the source given 
by~$aH\| \vec{x} \|\!=\!1$,
which coincides with the past particle horizon
 of a distant future observer at the origin.
We are interested in the effects within the light cone 
(nonshaded region) which capture the dynamical effects of graviton loops.
}
\label{conf diag}
\end{figure}

This section is devoted to solving Eq.~(\ref{point source correction}) 
for the one-loop graviton correction to the scalar point source potential.
We are interested in obtaining the solution at late times for which~$a\!\to\!\infty$,
after releasing the point source to interact with inflationary gravitons at 
the initial time~$\eta_0\!=\!-1/H$. We are interested in dynamical
corrections, which propagate within the future light cone of the source which --- from 
the  point of view of a late time local observer ---
encompasses both sub-Hubble, and super-Hubble distances away
from the point source, as illustrated
in Fig.~\ref{conf diag}.

First, the source on the right-hand side of~(\ref{point source correction}) 
needs to be computed, and we split it into seven parts,
\begin{equation}
\partial^2 \biggl[ \frac{\Phi_1(\eta,\| \vec{x} \|)}{ \| \vec{x} \| } \biggr]
	=  \sum_{n=1}^{7} K_n \, ,
\label{Kn split}
\end{equation}
according to the seven terms in the retarded self-mass~(\ref{retarded self mass}),
\begin{eqnarray}
&&
\hspace{-0.7cm}
K_1 =
	- \int \! d^4x' \, \partial^2 \partial'^2
	\biggl\{ \biggl[ \frac{ \ln(aa') }{ 96\pi^2 } \!-\! \alpha \biggr] \frac{\delta^4(x\!-\!x')}{aa'}
		 \biggr\} \frac{1}{ \| \vec{x}^{\,\prime} \| } \, ,
\label{K1}
\\
&&
\hspace{-0.7cm}
K_2 =
	- \int \! d^4x' \, H^2 \partial \!\cdot\! \partial'
	\biggl\{ \biggl[ \frac{ 19 \ln(aa') }{ 96\pi^2 } \!+\! \beta \biggr] \delta^4(x\!-\!x')
		 \biggr\} \frac{1}{ \| \vec{x}^{\,\prime} \| } \, ,
\label{K2}
\\
&&
\hspace{-0.7cm}
K_3 =
	\int \! d^4x' \, H^2 \vec{\nabla} \!\cdot\! \vec{\nabla}'
	\biggl\{ \biggl[ \frac{ 5 \ln(aa') }{ 16\pi^2 } \!+\! \gamma \biggr] \delta^4(x\!-\!x')
		 \biggr\} \frac{1}{ \| \vec{x}^{\,\prime} \| } \, ,
\label{K3}
\\
&&
\hspace{-0.7cm}
K_4 =
	\int\! d^4x' \, \delta H^4 (a')^2 \delta^4(x\!-\!x') \frac{1}{ \| \vec{x}^{\,\prime} \| } \, .
\label{K4}
\\
&&
\hspace{-0.7cm}
K_5 =
	\frac{-1}{768 \pi^3}\int\! d^4x' \, \partial^2 \partial'^2 \biggl\{
	\frac{1}{aa'} \partial^4 \biggl[ 
	\theta \bigl( \Delta\eta \!-\! \| \Delta\vec{x} \| \bigr)
\nonumber \\
&&	
\hspace{4.5cm}
	\times \biggl(
	\ln\Bigl[ \mu^2 \bigl( \Delta\eta^2 \!-\! \| \Delta\vec{x} \|^2 \bigr)\Bigr] \!-\! 1 \biggr) \biggr]
		\biggr\} \frac{1}{\| \vec{x}^{\,\prime} \|} \, , \qquad
\label{K5}
\\
&&
\hspace{-0.7cm}
K_6 =
	\frac{1}{768\pi^3} \int\! d^4x' \, H^2 \partial^4 \bigl(19\partial'^2 \!-\! 18\nabla'^2 \bigl)
	\biggl\{
	\theta \bigl( \Delta\eta\!-\! \| \Delta\vec{x} \| \bigr)
\nonumber \\
&&	
\hspace{4.5cm}
	\times \biggl( \ln\Bigl[ \mu^2 \bigl( \Delta\eta^2 \!-\! \| \Delta\vec{x} \|^2 \bigr) \Bigr] \!-\! 1 \biggr)
	\biggr\} \frac{1}{\| \vec{x}^{\,\prime} \|} \, , \qquad
\label{K6}
\\
&&
\hspace{-0.7cm}
K_7 =
	\frac{-1}{64\pi^3} \int\! d^4x' \, H^2 \partial^4 \nabla'^2
	\biggl\{
	\theta \bigl( \Delta\eta\!-\! \| \Delta\vec{x} \| \bigr)
\nonumber \\
&&	
\hspace{4.5cm}
	\times \biggl( \ln\Bigl[ \tfrac{1}{4} H^2 
		\bigl( \Delta\eta^2 \!-\! \| \Delta\vec{x} \|^2 \bigr) \Bigr] \!+\! 1 \biggr)
	\biggr\} \frac{1}{\| \vec{x}^{\,\prime} \|} \, . \qquad
\label{K7}
\end{eqnarray}
In the last two integrals we have used that the derivatives act on a
function of relative coordinates only to change some of them into primed ones 
for later convenience.
Evaluating the first four source integrals is straightforward,
\begin{eqnarray}
&&
K_1 =
	4\pi \partial^2 \biggl\{ \frac{ \delta^3(\vec{x}) }{ a^2 } 
		\biggl[ \frac{\ln(a)}{48\pi^2} \!-\! \alpha \biggr] \biggr\} \, ,
\label{K1sol}
\\
&&
K_2 =
	- 4\pi \delta^3(\vec{x}) H^2 \biggl[ \frac{19 \ln(a) }{ 48\pi^2 } \!+\! \beta \biggr] \, ,
\label{K2sol}
\\
&&
K_3 =
	4\pi \delta^3(\vec{x}) H^2 \biggl[ \frac{5 \ln(a) }{ 8\pi^2 } \!+\! \gamma \biggr] \, ,
\label{K3sol}
\\
&&
K_4 =
	\frac{\delta H^4 a^2}{ \| \vec{x} \| } \, .
\label{K4sol}
\end{eqnarray}
For the remaining three sources it proves best to first take all the 
unprimed derivatives out of the integral, then to integrate
by parts the remaining primed derivatives onto the classical point source
potential, and use the classical equation of motion,
\begin{equation}
\nabla^2 \frac{1}{\| \vec{x} \|} = \partial^2 \frac{1}{\| \vec{x} \|} = -4\pi \delta^3(\vec{x}) \, .
\end{equation}
This procedure is exact for integrating~$\nabla^2$ by parts, while for~$\partial'^2$
we drop the surface terms from the initial time surface, which decay at late times,
and can be absorbed into non-Gaussian corrections of the initial state~\cite{Kahya:2009sz}
(the integrals corresponding to the terms we drop were computed 
in~{\it e.g.} \cite{Glavan:2013jca}).
The delta function allows us to integrate over the spatial coordinates, leaving
single temporal integrals,
\begin{eqnarray}
&&
\hspace{-0.7cm}
K_5 =
	\frac{ \partial^2 }{192 \pi^2}
	\frac{1}{a} \partial^4 \!\! \! 
	\int\limits_{-1/H}^{\eta} \!\!\!  d\eta' \, 
	\frac{1}{a'} 
	\theta \bigl( \Delta\eta \!-\! \| \vec{x} \| \bigr)
	\, \biggl\{
	\ln\Bigl[ \mu^2 \bigl( \Delta\eta^2 \!-\! \| \vec{x} \|^2 \bigr)\Bigr] \!-\! 1 \biggr\} \, , 
\\
&&
\hspace{-0.7cm}
K_6 =
	- \frac{ H^2 \partial^4 }{192\pi^2} \!\!
	\int\limits_{-1/H}^{\eta} \!\!\!  d\eta' \, 
	\theta \bigl( \Delta\eta\!-\! \| \vec{x} \| \bigr)
	\, \biggl\{ \ln\Bigl[ \mu^2 \bigl( \Delta\eta^2 \!-\! \| \vec{x} \|^2 \bigr) \Bigr] \!-\! 1 \biggr\}
	\, , 
\\
&&
\hspace{-0.7cm}
K_7 =
	\frac{H^2 \partial^4 }{16\pi^2} \!\!
	\int\limits_{-1/H}^{\eta} \!\!\!  d\eta' \, 
	\theta \bigl( \Delta\eta\!-\! \| \vec{x} \| \bigr)
	\, \biggl\{ \ln\Bigl[ \tfrac{1}{4} H^2 
		\bigl( \Delta\eta^2 \!-\! \| \vec{x} \|^2 \bigr) \Bigr] \!+\! 1 \biggr\}
	  \, ,
\end{eqnarray}
which are all elementary, and evaluate to
\begin{eqnarray}
&&
\hspace{-0.7cm}
K_5 =
	\frac{ \partial^2 }{192 \pi^2}
	\frac{1}{a} \partial^4
	\biggl\{ \theta\bigl(\Delta\eta_0 \!-\! \| \vec{x} \|  \bigr)
		\biggl[ H \bigl( \Delta\eta_0^2 \!-\! \| \vec{x} \|^2 \bigr)
			\biggl( \frac{1}{2} 
				\ln\Bigl[ \mu^2 \bigl( \Delta\eta_0^2 \!-\! \| \vec{x} \|^2 \bigr) \Bigr]
					- 1 \biggr)  
\nonumber \\
&&
	+ \frac{1}{a} \biggl( - 2 \| \vec{x} \| \, \ln \bigl( 2\mu \| \vec{x} \| \bigr)
			- 3 \bigl( \Delta\eta_0 \!-\! \| \vec{x} \| \bigr)
		+ \bigl( \Delta\eta_0 \!-\! \| \vec{x} \| \bigr) 
			\ln\Bigl[ \mu \bigl( \Delta\eta_0 \!-\! \| \vec{x} \| \bigr) \Bigr]
\nonumber \\
&&
	\hspace{1cm}
	+ \bigl( \Delta\eta_0 \!+\! \| \vec{x} \| \bigr) 
			\ln\Bigl[ \mu \bigl( \Delta\eta_0 \!+\! \| \vec{x} \| \bigr) \Bigr] \biggr)
	\biggr] \biggr\} \, , 
\\
&&
\hspace{-0.7cm}
K_6 =
	- \frac{ H^2 \partial^4 }{192\pi^2} 
	\biggl\{ \theta\bigl(\Delta\eta_0 \!-\! \| \vec{x} \|  \bigr)
		\biggl[ - 2 \| \vec{x} \| \, \ln \bigl( 2\mu \| \vec{x} \| \bigr)
			- 3 \bigl( \Delta\eta_0 \!-\! \| \vec{x} \| \bigr)
\\
&&
	\hspace{0cm}
	+ \bigl( \Delta\eta_0 \!-\! \| \vec{x} \| \bigr) 
			\ln\Bigl[ \mu \bigl( \Delta\eta_0 \!-\! \| \vec{x} \| \bigr) \Bigr]
	+ \bigl( \Delta\eta_0 \!+\! \| \vec{x} \| \bigr) 
			\ln\Bigl[ \mu \bigl( \Delta\eta_0 \!+\! \| \vec{x} \| \bigr) \Bigr]
			 \biggr]
	 \biggr\}
	\, , 
\nonumber \\
&&
\hspace{-0.7cm}
K_7 =
	\frac{H^2 \partial^4 }{16\pi^2} 
	\biggl\{ \theta\bigl(\Delta\eta_0 \!-\! \| \vec{x} \|  \bigr)
		\biggl[ - 2 \| \vec{x} \| \, \ln \bigl( H \| \vec{x} \| \bigr)
			- \bigl( \Delta\eta_0 \!-\! \| \vec{x} \| \bigr)
\\
&&
	\hspace{0cm}
	+ \bigl( \Delta\eta_0 \!-\! \| \vec{x} \| \bigr) 
			\ln\Bigl[ \tfrac{1}{2} H \bigl( \Delta\eta_0 \!-\! \| \vec{x} \| \bigr) \Bigr]
	+ \bigl( \Delta\eta_0 \!+\! \| \vec{x} \| \bigr) 
			\ln\Bigl[ \tfrac{1}{2} H \bigl( \Delta\eta_0 \!+\! \| \vec{x} \| \bigr) \Bigr]
			 \biggr]
	 \biggr\}
	  \, .
\nonumber 
\end{eqnarray}
The final step in evaluating these is to act with all the external derivatives,
except for the one~$\partial^2$, which is useful to keep as is, since it allows us to
invert the equation of motion~(\ref{Kn split}) by simply dropping it. However, we 
must not forget that this~$\partial^2$ still acts on a function, and it annihilates its
homogeneous solutions, which yields rather simple results,
\begin{eqnarray}
&&
\hspace{-0.7cm}
K_5 =
	\frac{ \partial^2 }{48 \pi^2}
	\biggl[ \frac{\theta\bigl(\Delta\eta_0 \!-\! \| \vec{x} \|  \bigr)}{ \| \vec{x} \| } 
		\times \frac{1}{ (a \| \vec{x} \|)^2} \biggr] \, , 
\label{K5sol}
\\
&&
\hspace{-0.7cm}
K_6 =
	\frac{ H^2 \partial^2 }{48\pi^2} 
	\biggl[ \frac{ \theta\bigl(\Delta\eta_0 \!-\! \| \vec{x} \|  \bigr) }{ \| \vec{x} \| }
		\times \ln \bigl( 2\mu \| \vec{x} \| \bigr)
	 \biggr]
	\, , 
\label{K6sol}
\\
&&
\hspace{-0.7cm}
K_7 =
	\frac{H^2 \partial^2 }{4\pi^2} 
	\biggl[ \frac{ \theta\bigl(\Delta\eta_0 \!-\! \| \vec{x} \|  \bigr) }{ \| \vec{x} \| }
		\times \Bigl( - \ln \bigl( H \| \vec{x} \| \bigr) - 1 \Bigr)
	 \biggr]
	  \, . \qquad
\label{K7sol}
\end{eqnarray}
In the expression above we did not bother to keep the terms with support
only on the light cone, or outside of it, as in the late time limit the entire
region of physical interest is within the light cone of the point source
released to interact at~$\eta_0\!=\!-1/H$, as depicted in Fig~\ref{conf diag}.
In what follows we drop the theta function from the three sources above, 
and explicitly focus on corrections inside the light cone.

\medskip

Inverting equation~(\ref{Kn split}) for sources~(\ref{K1sol})--(\ref{K4sol}) 
and~(\ref{K5sol})--(\ref{K7sol}) we just computed yields the correction to the point source 
potential we are after. This is trivial for sources~$K_1$ and~$K_5$--$K_7$, as
it simply involves dropping the overall~$\partial^2$ from the sources.
Inverting sources~$K_2$--$K_4$ is only slightly more involved. It is facilitated
by noting the following two identities for d'Alembertian operators acting
on spherically symmetric functions:
\begin{eqnarray}
&&
\partial^2 \biggl[ \frac{f\bigl( \eta \!\mp\! \| \vec{x} \| \bigr)}{ \| \vec{x} \| } \biggr]
	= - 4 \pi \delta^3(\vec{x}) f( \eta) \, ,
\\
&&
\partial^2 \biggl[ \frac{f(\eta)}{\| \vec{x} \|} \biggr]
	= - 4 \pi \delta^3(\vec{x}) f(\eta)
		- \frac{\partial_0^2 f(\eta)}{\| \vec{x} \|} \, .
\end{eqnarray}
These are easily proven by specializing the d'Alembertian operator
to functions of just $\eta$ and $\Vert \vec{x}\Vert$ and then factorizing it,
\begin{equation}
\partial^2 = - \frac{1}{\| \vec{x} \|}
	\biggl[ \partial_0 \!-\! \frac{\partial}{\partial \| \vec{x} \|} \biggr]
	\biggl[ \partial_0 \!+\! \frac{\partial}{\partial \| \vec{x} \|} \biggr]
	\| \vec{x} \| \, .
\end{equation}
The inversion for sources~$K_2$--$K_4$ involves two particular identities,
\begin{eqnarray}
&&
\partial^2 \biggl\{ \frac{ 1 }{ \| \vec{x} \| }
	\ln\Bigl[ H \bigl( \| \vec{x} \| \!-\! \eta \bigr) \Bigr] \biggr\}
	=
	4\pi \delta^3(\vec{x}) \ln(a) \, ,
\\
&&
\partial^2 \biggl\{ 
	\frac{\ln(a)}{\| \vec{x} \|}
	+ \frac{ 1 }{ \| \vec{x} \| }
	\ln\Bigl[ H \bigl( \| \vec{x} \| \!-\! \eta \bigr) \Bigr] \biggr\}
	=
	- \frac{H^2a^2}{\| \vec{x} \|} \, .
\qquad
\end{eqnarray}
This determines the graviton one-loop correction to the point-source potential
at late times,
\begin{eqnarray}
&&
\hspace{-0.7cm}
\Phi_1 \bigl( \eta, \| \vec{x} \| \bigr)
	=
	4\pi \biggl[ \frac{\ln(a)}{48\pi^2} \!-\! \alpha \biggr] 
		\bigl( a \| \vec{x} \| \bigr) \delta^3(a\vec{x})
	+ \frac{1}{48 \pi^2 (a\| \vec{x} \|)^2}
\nonumber \\
&&
\hspace{2cm}
	+ \frac{H^2}{48\pi^2} \biggl[ 
		- 48 \pi^2 \delta \ln\bigl( 1 \!+\! aH\| \vec{x} \| \bigr)
		+ 11 \ln \Bigl( \frac{1}{a} \!+\! H\| \vec{x} \| \Bigr)
\label{point source solution}
\\
&&
\hspace{4cm}
		+ \ln \bigl( 2\mu \| \vec{x} \| \bigr) 
		- 12 \ln\bigl( H\| \vec{x} \| \bigr) - 12
		+ 48\pi^2 (\beta \!-\! \gamma)
		\biggr] \, .
\nonumber 
\end{eqnarray}
We have determined this one-loop graviton correction to the point source potential 
up to homogeneous terms. However, these 
necessarily take the form of surface terms from the initial
time surface, and thus can be absorbed into 
perturbative non-Gaussian initial state corrections~\cite{Kahya:2009sz}.
Our result captures the dynamical effects 
generated by interactions that do not depend on the choice of the initial state.

\section{Discussion and conclusions}
\label{Discussion and conclusions}

In this work we have investigated graviton loop corrections 
to a massless, conformally coupled scalar on a de Sitter background,
with a particular emphasis on large logarithms whose gauge dependence
could be the object of further study. Our main results are the plane
wave scalar mode function (\ref{f result}) and the exchange potential
(\ref{point source solution}). We discuss each in turn.

\bigskip

\noindent{\bf Dynamical scalar corrections.}
The late-time limit of a plane wave is
\begin{eqnarray}
&&
\phi(\eta,\vec x)
	= \phi_0(\eta,\vec x)
	\Biggl\{1\!+\! \hbar GH^2\biggl[-16\pi\delta\ln(a)
\nonumber\\
&&\hspace{5cm}
		+\, \frac{ik}{H}\left(\!32\pi\delta
              \!+\!\frac{19}{3\pi}\right)\!\frac{\ln(a)}{a} 
		+{\rm const.}\biggr] 
		\Biggr\}
\,,
\qquad
\label{u_1 relevant}
\end{eqnarray}
where 
$\phi_0(\eta,\vec x) \!=\! e^{-ik\eta+i\vec k\cdot\vec x}/a$
is the tree-level contribution, $G$ is Newton's constant,
 and we have restored the reduced Planck constant~$\hbar$. The 
large logarithm in (\ref{u_1 relevant}) vanishes if we choose
the $R^2 \phi^2$ counterterm $\delta \!=\! 0$. The decaying 
logarithm $\ln(a)/a$ comes from the local part of the retarded 
self-mass-squared~(\ref{retarded self mass}), while the constant 
contribution originates from both the local and the nonlocal parts.
The constant contribution also depends on the choice of the initial 
state and for that reason cannot be fixed. The decaying logarithm
does cause the time derivative of the conformally rescaled field to
grow relative to its classical value, and that might be significant
\cite{Friedrich:2019hev}.

We should also comment on the work of Boran, Kahya and Park who studied 
the same system \cite{Boran:2014xpa,Boran:2017fsx,Boran:2017cfj}. Their 
result for the self-mass  
was given in
Refs.~\cite{Boran:2014xpa,Boran:2017fsx}, while their solution for scalar
plane waves appears in equations (44) and (56) of
Ref.~\cite{Boran:2017cfj}. 
Their leading one-loop corrections are of order $a \ln(a)$ and $a$, and are claimed to 
originate from the nonlocal contributions. In contrast, the only nonlocal
contributions we find come from the lower limits of temporal integrations 
and fall off at late time. They also claim a $\ln(a)$ enhancement from the 
local part of the self-mass~(\ref{L ctm}) as we do, but they get it 
from the coupling constant $\gamma$ (their $-\Delta c_4$), whereas ours
comes from $\delta$ (related to their $\Delta c_3$). We are unable to 
account for these discrepancies but it might be relevant to note that
they employed a cumbersome de Sitter invariant representation in which 
surface terms must be handled with great care \cite{Leonard:2012si}.
Fr\"ob also reported a problem with the flat space correspondence limit 
of their result \cite{Frob:2017apy}.

\bigskip

\noindent{\bf Point source corrections.}
At late times the one-loop corrected exchange potential is given by
Eqs.~(\ref{pt source pert expansion}) and~(\ref{point source solution}),
\begin{eqnarray}
&&
\hspace{-0.7cm}
\phi(\eta,\vec{x}) = \frac{-1}{4\pi a r}
	\Biggl\{
	1 
	+ \frac{\hbar G}{3\pi (a r)^2 } 
	+ \frac{4\hbar G}{3} 
		\Bigl( \ln(a) \!-\! 48\pi^2 \alpha \Bigr) \bigl( a r \bigr) \delta^3(a\vec{x})
\nonumber \\
&&	\hspace{2cm}
	+ \frac{\hbar G H^2}{3\pi} \biggl[ - 48 \pi^2 \delta \ln\bigl( 1 \!+\! aH r \bigr)
	+ 11 \ln\biggl( \frac{1 \!+\! aH r }{ aH r } \biggr)
\nonumber \\
&&	\hspace{4cm}
	- \ln\Bigl( \frac{\hbar H}{2\mu}\Bigr) 
	- 12 + 48\pi^2 (\beta\!-\!\gamma) \biggr]
	\Biggr\} \, ,
\label{pt source final}
\end{eqnarray}
where $r \!\equiv\! \| \vec{x}\|$. 
This result captures corrections from graviton loops inside the light cone of the 
point source, as depicted by the white region in Fig.~\ref{conf diag}.
Note that the constant terms in the last line of the result above 
contain a part that is logarithmically dependent on the arbitrary 
renormalization scale~$\mu$. This term can be reinterpreted as a logarithmic 
running of the coupling constants $\beta \!-\!\gamma$ from Eq.~(3), and could be
used to cancel all the constant terms.\footnote{
The running of $\beta \!-\! \gamma$ with $\mu$ is determined by the beta function, 
$\beta_{\beta-\gamma} \!=\! 1/(48\pi^2)$.
}

There are two interesting regimes of (\ref{pt source final})---the  sub-Hubble 
regime of $a r \!\ll\! 1/H$ and the super-Hubble regime of $a r \!\gg\! 1/H$.
In the sub-Hubble regime the potential reduces to
\begin{eqnarray}
&&
\hspace{-0.7cm}
\phi(t,\vec x) 
\xrightarrow{aH r \ll 1}
16\pi\hbar G
\biggl[\alpha\!-\!\frac{\ln(a)}{48\pi^2}\biggr]
\delta^3(a\vec{x})
\!-\!\frac{1}{4\pi a r}
\Biggl\{1 + \frac{\hbar G}{3\pi a^2 r^2}
\qquad
\nonumber
\\
&&	\hspace{4cm}
	+\, \frac{\hbar GH^2}{3\pi} \biggl[ -11 \ln\bigl(aH r \bigr) 
	+ {\rm irrelevant}
\biggr]
\Bigg\} \, . \qquad
\label{sH limit}
\end{eqnarray}
The delta function contribution arises from the first 
term in~(\ref{L ctm}), and the secular correction
$\propto\ln(a)$ acts as a dynamical screening of $\alpha$.
The flat space limit~$a\!\to\!1$ and~$H\!\to\!0$ is captured by the terms in 
the first line of~(\ref{sH limit}), which contains only conformally rescaled 
flat space corrections. The second line in~(\ref{sH limit}) is of a purely de 
Sitter origin and contains a large logarithm and a constant term. The logarithm
can be seen as a logarithmic antiscreening of the source. However, its effect 
is small compared with the conformally rescaled flat space correction.

\medskip

In the super-Hubble regime the potential~(\ref{pt source final}) reduces to
\begin{eqnarray}
\hspace{-0.cm}
\phi(\eta,\vec{x}) \xrightarrow{aH r \gg 1} \frac{-1}{4\pi a r}
	\biggl\{
	\!1 
	\!+\! 16\pi\hbar G H^2\Bigl[-\delta \ln\bigl( aH r \bigr)
	\!+\! {\rm irrelevant}
\Bigr]
	\biggr\} 
. \;\;
\label{SH limit}
\end{eqnarray}
The large logarithm can be eliminated by choosing $\delta = 0$,
which also eliminates the large logarithm in the scalar plain wave.
It therefore seems that the massless, conformal scalar is not a 
good venue for studying the gauge dependence of large logarithms 
from inflationary gravitons. 

\section*{Acknowledgements}

This work was partially supported 
by the Fonds de la Recherche Scientifique-- FNRS under Grant IISN 4.4517.08 -- 
	Theory of fundamental interactions; 
by Taiwan MOST grant 108-2112-M-006-004; 
by the D-ITP consortium, a program of the Netherlands Organization for Scientific 
	Research (NWO) that is funded by the Dutch Ministry of Education, Culture 
	and Science (OCW); 
by NSF grants PHY-1806218 and PHY-1912484; 
and by the Institute for Fundamental Theory at the University of Florida. 


\end{document}